# An equation of state for low and high energy Bose-Einstein condensation[1]

**V Barbarani**

Member of European Physical Society, via delle Panche 140, 51100 Florence, Italy

E-mail: *v.barbarani@infogroup.it*

**Abstract.** The aim of this work is to investigate how energy depends on the two-body interaction potential in Bose-Einstein condensation (BEC) phenomena. An equation of state is obtained which is valid both for low and high energy BEC, through the application of a revised form of quantum statistics. An extension of the singularity conditions describing the state of BEC is given, in order to consider interactions between particles due to a central interatomic potential. From the singularity conditions of the corresponding system of hard-sphere bosons and the equation for the energy of the system in its ground state, the equation of state connecting temperature, density and energy in BEC is deduced, with upper and lower limits for the energy depending on the form of the central interaction potential. It is shown that high energy mode is allowed in the case of Coulomb type interaction only, low energy mode in the case of non-Coulomb type interaction. Numerical results are then derived for low energy BEC, occurring in neutral matter, with application to He-4 and alkali-metal atoms, and in the case of high energy BEC occurring in systems of charged bosons, with application to atomic nuclei.

## 1. Introduction

The theoretical work following the pioneering experiments with sodium, rubidium and lithium [1-3], has enlightened the role played by two-body collisions in Bose-Einstein condensation (BEC), above all as far as dilute gas of neutral atoms is concerned in connection with mean field theory and Gross-Pitaevskii equation (see [4] for a summary of the quantitative aspects of the theory, while a qualitative treatment of the process is reported in [5]). At the same time the opinion that Bose Einstein condensation, is to be considered as a unified paradigm to interpret many phenomena in condensed matter, atomic and nuclear physics emerged and is now widely diffused [6]. The idea to consider the state of matter in the nucleus as Bose-Einstein condensation was suggested since 1998 by Kim and Zubarev [7]. Successively the BEC mechanism has been applied to the study of a system of identical charged bosons ("Bose" nuclei) confined in ion traps, by solving approximately many-body Schrodinger equation, in order to explain low energy nuclear reactions in matter [8, 9]. More recently the existence of Bose-Einstein condensate of weakly interacting $\alpha$ particles in finite nuclei, has been proposed [10] and it is now intensely investigated [11, 12]. This condensate of $\alpha$ particles in nuclear matter, is considered to be an analogue to Bose-Einstein condensate for finite number of dilute bosonic atoms at very low temperature, (see [13] and references therein for a comprehensive discussion of this approach). A question arises about the differences between the energy range of BEC processes in neutral atoms, which occur at extremely low energies with temperature T ≈ 0 K, and in nuclear matter, where charged particles are to be considered and the value of energy is in the order of MeV.

---

[1] A preliminary version (extended abstract) of the work which is treated here in full details, was accepted for poster contribution at the "22nd General Conference of the Condensed Matter Division of the European Physical Society" (CMD22, August 25th-29th, 2008, Rome).

The aim of the statistical theory we propose in this paper, is to give a unified treatment of two-body interactions occurring in BEC, which is valid for all density values. As a result of the theory a general relation between temperature and density is obtained and the wide range in the energies implied by BEC processes is explained simply. The theory is developed within the context of uniform Bose gases [14], with no confining potential applied: a new form of quantum statistics, the author has recently proposed [15] for the ideal gas, is adopted and modified in order to include interactions between bosons. The new quantum statistics for bosons and fermions, we denote respectively as BE$\omega$ and FD$\omega$, to distinguish them from the classical quantum statistics of Bose-Einstein (BE) and Fermi-Dirac (FD), are obtained by the minimization of the energy of microscopic configurations of the system. While distributions BE and FD are functions of level degeneracy $g(\varepsilon)$, therefore of all accessible quantum states with energy $\varepsilon$, new distributions BE$\omega$ and FD$\omega$ are functions of $\omega(\varepsilon)$, the number of quantum states effectively occupied by particles at $\varepsilon$ energy. It is to be pointed out that the new form of quantum statistics we adopt here, does not mean that the classical ones are wrong: as it is clearly stated in [15], from the assumption of the equivalence between the two forms of quantum statistics, a simple and general formulation of quantum coherence is obtained. A unified definition of BEC phenomena is therefore allowed by the new statistics, both in the case of low density gases and in the case of degeneracy, as superfluid $^4He$ and $^3He$. The new statistics defines BEC as a single state of quantum coherence and gives exact physical conditions in order that such a state can exist.

In order to include interactions between particles, we consider an interatomic potential of central type $\propto 1/r^\gamma$, that is depending only on the distance $r$ between two particles, with positive exponent $\gamma$, and then we get the singularity conditions of quantum states from BE$\omega$ statistics, in the more realistic case of a gas with interacting bosons. The mathematical procedure described above is equivalent to formulate a quantum statistics for hard-sphere bosons and it leads to an exact definition of the bosons' diameter $a(\varepsilon)$, as a function of the energy $\varepsilon$ in the single state.

As far as hard-sphere models are concerned in studying fluids consisting of interacting particles, there exists a wide literature dealing with the applications of such models by analytical and numerical methods, starting since 1950s [16-20]. In this work we use the results of Oden, Henderson and Coleman [21], in applying the model first proposed by Barker [22] in the case of classical fluids, to the study of quantum fluids. The simple energy equation obtained from this model for a system of rigid spheres in their ground state, and the equation for the single state of quantum coherence obtained from BE$\omega$ statistics, allow us to derive the theoretical relation among temperature, energy and density in a system of bosons in BEC. From this analysis a further important result is obtained, as to the way energy varies in BEC processes. From the new form of quantum statistics we get an infinite number of accessible single states of quantum coherence, in the case of the ideal gas, both at low and high energy. Including interactions in BE$\omega$ statistics, we can prove that the "direction" of the energy of single states, depends on the value of exponent $\gamma$ of interatomic potential: if $\gamma > 2$ BEC can occur at "low" energy only; on the contrary if $\gamma = 1$, corresponding with the Coulomb potential, only the high energy mode is allowed. In the following we demonstrate that in the case of interactions between neutral atoms the value $\gamma = 5$ can be assumed. Obviously in the case of interactions between charged bosons, like those occurring in nuclear processes, the right value is $\gamma = 1$.

The work is organized as follows.

In Section 2 a brief summary of the main results of BE$\omega$ statistics in the case of the ideal gas is given. Then the quantum statistics for hard-sphere bosons is developed, introducing interactions in BE$\omega$ statistics, and the corresponding singularity conditions determining the physical state of BEC, are derived.

In Section 3 the general form of the equation of state for a system of *N* identical bosons in BEC is deduced for any value of $\gamma$, through the assumptions the system is in thermal equilibrium and in its ground state.

In Section 4 a central interatomic potential, depending on atomic dimensions, is proposed for neutral atoms with s-type outermost orbitals. This is the case of $^4He$ and alkali-metal atoms used in BEC experiments. The spherical shape of s-type orbitals and the assumption of purely electrostatic forces, are used to derive the interaction potential. The numerical results obtained from the theory, are then used to plot the temperature-density relation for these atoms in BEC.

In Section 5 the generalized theory proposed in Section 2 and 3 is applied to the study of the condensation into a single state of quantum coherence, of a system of identical charged bosons. In particular the condensation of deuterons D ($^2$H nuclei) is considered and the equation of state is drawn for the whole range of admissible energy and density. The region of ordinary nuclear matter is then considered in details and the connection between the energy $\varepsilon$ of particles in BEC, and the binding energy is studied.

Summary and conclusions are reported in Section 6.

## 2. BE $\omega$ statistics for hard-sphere bosons

In that follows we give a brief summary of the new quantum statistics in the case of the ideal Bose gas. A complete treatment of the subject is found in [15]. Then the statistics is extended to the case of an interacting gas of hard-sphere bosons. As to Bose-Einstein statistics, following the classical treatment in [23], we will use the continuous approximation with degeneracy parameter *A* (instead of chemical potential $\mu$ used in modern notation, $A = e^{\mu/kT}$, *k* is the Boltzmann constant).

Let us consider a system in thermal equilibrium made up of *N* identical particles (bosons) with mass *m*. Classical BE distribution establishes that the number *n* of particles in the same energy level $\varepsilon$ (for the sake of brevity in the following we shall use this expression instead of "with energy between $\varepsilon$ and $\varepsilon + d\varepsilon$") given by

$$n(\varepsilon) = \frac{Ag(\varepsilon)}{e^{\beta\varepsilon} - A} \qquad (1)$$

with $0 < A \leq 1$, $g(\varepsilon)$ is the distribution function of quantum states per unit volume (or the density of accessible quantum states with energy $\varepsilon$), and $\beta$ depends on temperature through the formula

$$\beta = \frac{1}{kT} \ .$$

As to the range of possible energy values, all the values in the interval $0 \leq \varepsilon \leq \infty$ are allowed by BE statistics and constant *A* and number density $n_\delta$ are then connected through the equation (assuming a fixed number of particles):

$$\int_0^\infty \frac{Ag(\varepsilon)}{e^{\beta\varepsilon} - A} d\varepsilon = n_\delta . \qquad (2)$$

From the alternative formulation of BE $\omega$ quantum statistics the following equation holds too:

$$n(\varepsilon) - 1 = \frac{\omega^2(\varepsilon)}{e^{\beta\varepsilon} - 1} \qquad (3)$$

where the function $\omega(\varepsilon)$ is defined as the number of quantum states with energy $\varepsilon$ which are occupied at least by one particle. In other words the statistics takes into consideration only the occupied quantum states, which give some contribution to the energy of the ensemble (obviously all other states have $n(\varepsilon) = 0$ since they are not occupied). Therefore it is $\omega(\varepsilon) \geq 1$ by definition.

Assuming the equivalence of BE and BE$\omega$ statistics, since both of them give the same distribution of particles of a system as function of energy, and from the existence condition $\omega^2(\varepsilon) > 0$, it is easy to prove that the admissible energy interval is finite $\varepsilon_m \leq \varepsilon \leq \varepsilon_M$, with minimum and maximum values given by the intersection points between the function $g(\varepsilon)$, which we know to have the form

$$g(\varepsilon) = b\sqrt{\varepsilon} \qquad (4)$$

and the exponential function

$$f(\varepsilon, T, A) = \frac{e^{\beta\varepsilon} - A}{A} . \qquad (5)$$

The parameter $A$ measures the level of degeneration and when $A = 1$ we say the fluid is in complete degeneracy conditions. The constant $b$ in (4) is

$$b = \frac{4\pi}{h^3} M (2m^3)^{1/2}$$

where the weight factor $M$ is equal to the total number of possible projections of particle angular momentum [24].

According to the theory of quantum statistics BE$\omega$, the Bose-Einstein condensation corresponds with a single state of quantum coherence, that is to say all particles occupy the same quantum state with energy $\varepsilon$. Such a state is defined by the triplet $(\varepsilon, T, A)$ which satisfies the conditions

$$g(\varepsilon) = f(\varepsilon, T, A) \qquad (6)$$

$$\frac{\partial g(\varepsilon)}{\partial \varepsilon} = \frac{\partial f(\varepsilon, T, A)}{\partial \varepsilon} \qquad (7)$$

corresponding with geometrical tangency of curves $g(\varepsilon)$ and $f(\varepsilon, T, A)$ given in (4), (5). In the case of bosons the above equations become

$$\frac{1}{A} e^{\beta\varepsilon} - 1 = b\sqrt{\varepsilon} \qquad (8)$$

$$\frac{\beta}{A} e^{\beta\varepsilon} = \frac{1}{2} \frac{b}{\sqrt{\varepsilon}} \qquad (9)$$

from which the relation between energy and temperature in BEC follows

$$\sqrt{\varepsilon} = \frac{1}{2b}\left(\sqrt{1 + 2kTb^2} - 1\right)$$

or

$$\frac{kT}{2\varepsilon} \cong 1 \qquad (10)$$

since $1/(2b) \ll 1$. The ratio $\rho = kT/2\varepsilon$ between temperature and energy, has a fundamental role in the physics of BEC in the case of interacting gas too. Analogous formulas holds also in the case of fermions, except for the key role which is played by the Pauli exclusion principle and by the statistical conditions controlling its validity, as reported in [15].

In the treatment of quantum statistics given above, the interatomic interaction potential is assumed to be zero. To study the dependence of temperature on density in BEC, it is necessary to take into consideration the interatomic interaction energy too. In order to derive the equations for the single state of quantum coherence in the case of interacting gas, we assume a central two-body interaction potential of the form

$$\varphi(r) = \frac{\Phi}{r^{\gamma}} \qquad (11)$$

with positive $\gamma$ and constant $\Phi$. This implies a simple modification of function $g(\varepsilon)$, density of quantum states, just because a particle with energy $\varepsilon$ interacting with another particle, cannot get over the potential wall with radius $r$ such that

$$\varphi(r) = \frac{\Phi}{r^{\gamma}} > \varepsilon$$

or, in other words, it cannot draw nearer than $\left(\frac{\Phi}{\varepsilon}\right)^{\frac{1}{\gamma}}$. Therefore the "sphere" with volume

$$v = \frac{4}{3}\pi \left(\frac{\Phi}{\varepsilon}\right)^{\frac{3}{\gamma}}$$

around every particle is inaccessible for other particles with energy $\varepsilon$ or lesser. Hence if $N$ is the total number of atoms in the system and $V$ its volume, the resulting volume accessible to quantum states with energy $\varepsilon$ is given by

$$V - N\frac{4}{3}\pi\left(\frac{\Phi}{\varepsilon}\right)^{\frac{3}{\gamma}} = V\left[1 - \frac{4}{3}\pi \cdot n_{\delta}\left(\frac{\Phi}{\varepsilon}\right)^{\frac{3}{\gamma}}\right]$$

and the density $g(\varepsilon)$ becomes

$$g(\varepsilon) = \left[1 - \frac{4}{3}\pi \cdot n_{\delta}\left(\frac{\Phi}{\varepsilon}\right)^{\frac{3}{\gamma}}\right] b\sqrt{\varepsilon}. \qquad (12)$$

Since $g(\varepsilon) > 0$, from (12) we get a necessary and sufficient condition in order that accessibile quantum states can exist

$$\varepsilon > \Phi(\frac{4}{3}\pi \cdot n_{\delta})^{\frac{\gamma}{3}}. \qquad (13)$$

The central potential hypothesis (11) therefore is equivalent to consider the particles of the system as interacting hard-sphere bosons, with diameter $a(\varepsilon)$ depending on energy as follows

$$a(\varepsilon) = \left(\frac{\Phi}{\varepsilon}\right)^{\frac{1}{\gamma}}. \qquad (14)$$

In the quantum statistical theory we propose in this work, the quantity $a(\varepsilon)$ defined in (14) is used instead of the s-wave scattering length, usually assumed in current theories as diameter of hard-sphere bosons.

Let us now apply BE$\omega$ statistics to this system and find the BEC conditions, through the equations for the single state of quantum coherence (6) and (7), where $g(\varepsilon)$ is now defined by (12) and $f(\varepsilon, T, A)$ is the same function (5) as in the case of non-interacting gas. These conditions are represented in figure 1 Note the shift in the energy of function $g$, in the case of interacting gas, due to condition (13).

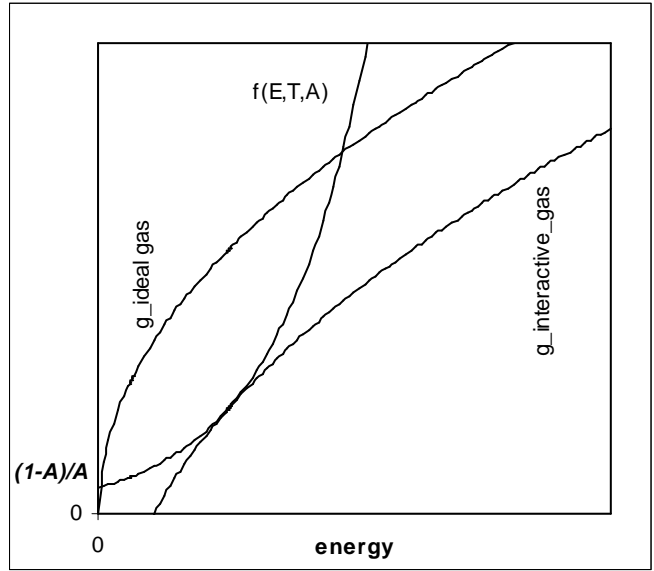

Figure 1 – The functions *f* and *g*, for the ideal and the interacting gas. For this one the geometrical conditions defining the state of BEC, according to BE$\omega$ theory, is represented (arbitrary units).

First of all we have to note that the conclusions about the admissible energy interval for quantum states remain unchanged: it is finite, $\varepsilon_m \leq \varepsilon \leq \varepsilon_M$, with minimum and maximum values given by the intersection points between the functions (5) and (12) defined above. As to the singularity conditions, instead of (8) and (9) we get

$$b\sqrt{\varepsilon}\,(1-\frac{4}{3}\pi\,n_\delta a^3(\varepsilon)) = \frac{1}{A}e^{\frac{\varepsilon}{kT}} - 1 \qquad (15a)$$

$$\frac{1}{2}\frac{b}{\sqrt{\varepsilon}}[1+(\frac{6}{\gamma}-1)\frac{4}{3}\pi\,n_\delta a^3(\varepsilon)] = \frac{1}{AkT}e^{\frac{\varepsilon}{kT}}. \qquad (15b)$$

Let us define

$$\frac{4}{3}\pi\,n_\delta a^3(\varepsilon) = n_0 \qquad (16)$$

$$(\frac{6}{\gamma}-1) = w \qquad (17)$$

$$\rho = \frac{kT}{2\varepsilon} \qquad (18)$$

then by taking

$$\frac{1}{A}e^{\frac{\varepsilon}{kT}} = \frac{kT}{2}\frac{b}{\sqrt{\varepsilon}}[1+(\frac{6}{\gamma}-1)\frac{4}{3}\pi\,n_\delta a^3(\varepsilon)]$$

from (15b) and substituting it in (15a) we get as to the ratio between temperature and energy

$$\rho = \frac{1-n_0}{1+wn_0} + \frac{1}{b\sqrt{\varepsilon}(1+wn_0)}. \qquad (19)$$

From (19) it follows (assuming density remains finite)

$$\lim_{\varepsilon \to \infty} \rho = 1 \qquad (20)$$

as in the case of the ideal gas (see (10)). Therefore at high energy the physical conditions of BEC in the real gas and in the ideal gas becomes similar. Since in all cases of physical interest it is $1/b\sqrt{\varepsilon}(1+wn_0) << 1$, from (19) we can assume

$$\rho = \frac{1-n_0}{1+wn_0} \qquad (21)$$

and the corresponding inverse equation

$$n_0 = \frac{1-\rho}{1+w\rho} \ . \qquad (22)$$

Equation (21) (and its companion (22)), which follows from the singularity conditions (15a-b), is the fundamental equation governing the physical parameters (temperature, density and energy) of a system of free particles in BEC. Note that the quantity $n_0$ defined in (16), measures the volume fraction occupied by interacting particles and the following upper limit holds by definition: $n_0 \leq 1$ (which is the same as (13)). The value of $\rho$ controls the level of dilution of the fluid: if $\rho \approx 1$, then $n_0 \approx 0$, which is the same as the condition $n_\delta a^3 \ll 1$, and the dilute gas (or weakly interacting) hypothesis [4] holds; on the contrary if $\rho \approx 0$, then $n_0 \approx 1$, and the fluid is a high density fluid. The statistical theory then applies both to dilute gases and high density fluids.

From (21) it follows that the temperature of the condensate is $T = 0$ when $n_0 = 1$, even if the quantum energy $\varepsilon$ of particles is not zero. This is because $T$ is the temperature of hard spheres with diameter $a$, hence it is a measure of their relative movement or shaking (see figure 2).

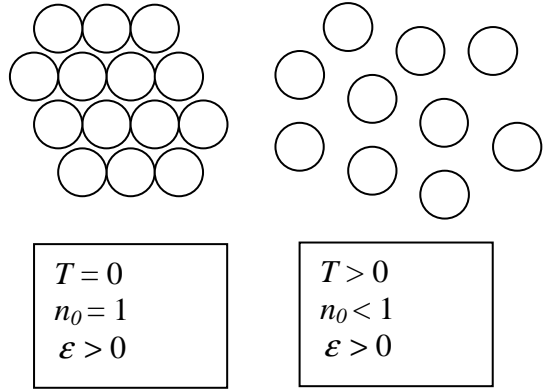

Figure 2 – A geometrical representation of the connection between temperature $T$, volume fraction $n_0$ occupied by hard spheres, and quantum energy $\varepsilon$ in BEC.

### 3. The general form of the equation of state for a system of hard-sphere bosons in its ground state

In order to derive the equation of state of the system in BEC, together with (21) we need an independent expression for the energy of bosons. This is obtained assuming the system is in its ground state. Many methods have been developed to handle the problem of many-body quantum system, where interacting particles are considered as hard-sphere bosons, which are based both on numerical solution of the Schrodinger equation, with the equivalent set of boundary conditions, and on theoretical treatments. As to the former approach an early work on the subject can be found in [18], as to the latter one see [19, 20]. Up-to-date references about the problem of the ground energy of a quantum system of hard-sphere bosons, can be found in [25] and in the already mentioned reference [8] for the case of charged bosons.

In that follows we use the results given in [21], in conjuction with equations (14) and (21) derived from BE $\omega$ statistics. The main reason why we use this model is it gives a simple expression for the ground energy of hard sphere bosons, due to the hypothesis of separation of the motion of particles into two components (the so-called tunnel model first proposed by Barker [22]). A brief comparison of the results obtained by this model with those derived by other methods, is reported in [26].

We can resume the basic assumptions of the model (see [26, 21] and references therein) as:
- a fluid is regarded as being composed of rows of molecules, which move along lines (or "tunnels"), which are bounded by neighboring rows of molecules;
- the motions of the molecules along and across the tunnels are separable and, to a good approximation, one-dimensional.
Then the energy of a particle in the ground state is given by two separate components

$$\varepsilon_2 = \frac{0.586 h^2}{8m(r-a)^2} \qquad (23)$$

and

$$\varepsilon_1 = \frac{h^2}{24m(l-a)^2} . \qquad (24)$$

The term $\varepsilon_2$ is the lowest energy level of a particle in a two-dimensional spherical square well of radius (*r-a*), and it corresponds with the transverse motion of the particle across the plane section of the tunnel; the term $\varepsilon_1$ is due to the one-dimensional motion of the particle in the tunnel. As for the geometrical dimensions, *a* is the diameter of the hard-spere bosons, *r* is the distance between the centers of neighboring rows and *l* is the mean spacing between particles along the same line. Density is connected with *r* and *l* through the relation

$$n_\delta = \frac{2}{\sqrt{3}} \frac{1}{lr^2}$$

which becomes

$$n_\delta = \frac{2}{\sqrt{3}} \frac{1}{r^3} \qquad (25)$$

since in order that the energy of the system is minimum, *r* and *l* are virtually equal. From (25) by simple algebraic manipulations we get

$$\varepsilon_2 = \frac{0.586 h^2}{8ma^2} \frac{1}{\left[\left(\frac{8\pi}{3\sqrt{3}}\frac{1}{n_0}\right)^{\frac{1}{3}} - 1\right]^2} \qquad (26)$$

$$\varepsilon_1 = \frac{h^2}{24ma^2} \frac{1}{\left[\left(\frac{8\pi}{3\sqrt{3}}\frac{1}{n_0}\right)^{\frac{1}{3}} - 1\right]^2} \qquad (27)$$

with $n_0$ defined in (16).
Depending on the way the particle moves in the tunnel, or the velocity modes, the energy can assume two distinct values:

$$\varepsilon = \varepsilon_2 + \varepsilon_1 \qquad (28)$$

in the case of three-dimensional motion, across the plane section and along the walls of the tunnel;

$$\varepsilon = \varepsilon_1 \qquad (29)$$

in the case of one-dimensional motion, along the walls of the tunnel only.
We can then assume the following general expression for $\varepsilon$

$$\varepsilon = \frac{Ch^2}{ma^2} \frac{1}{\left[\left(\frac{8\pi}{3\sqrt{3}} \frac{1}{n_0}\right)^{\frac{1}{3}} - 1\right]^2} \tag{30}$$

where $C$ is equal to

$$C = C2 = 0.1149 \tag{31}$$

in the case (28) or

$$C = C1 = \frac{1}{24} \tag{32}$$

in the case (29).

A general relation between temperature and density in BEC through energy $\varepsilon$, can now be derived from the theory developed above. From (30) and (14) remembering the definition of $n_0$ in (16), we get

$$n_\delta = \frac{2}{\sqrt{3}} \left(\frac{\varepsilon}{\Phi}\right)^{\frac{3}{\gamma}} \frac{1}{\left[h\left(\frac{\varepsilon}{\Phi}\right)^{\frac{1}{\gamma}} \sqrt{\frac{C}{m\varepsilon}} + 1\right]^3} \tag{33}$$

and by substituting this in (21)

$$T = \frac{2\varepsilon}{k} \frac{\left[h\left(\frac{\varepsilon}{\Phi}\right)^{\frac{1}{\gamma}} \sqrt{\frac{C}{m\varepsilon}} + 1\right]^3 - \frac{8\pi}{3\sqrt{3}}}{\left[h\left(\frac{\varepsilon}{\Phi}\right)^{\frac{1}{\gamma}} \sqrt{\frac{C}{m\varepsilon}} + 1\right]^3 + w\frac{8\pi}{3\sqrt{3}}} . \tag{34}$$

In Section 4 and 5 the value of $\Phi$ will be determined in the case of BEC of neutral alkali atoms and the case of charged bosons $^2$H (deuterons) and a graph of the function $T(n_\delta)$ is given.

Another important theoretical result can be deduced from the singularity conditions of BE$\omega$ statistics applied to a system of hard-sphere bosons in its ground state. New quantum statistics allow an infinite number of single states of quantum coherence to exist, which are solutions of (8) and (9), both at low and high energy. On the contrary the theory developed before for interacting gas, discriminates between low and high energy states depending on the exponent $\gamma$ of central potential (11). To see this let us rewrite the condition (13) as

$$n_\delta < \frac{3}{4\pi}\left(\frac{\varepsilon}{\Phi}\right)^{\frac{3}{\gamma}}$$

hence, by considering the above limitation in (33) we get

$$\varepsilon^{\frac{2-\gamma}{2\gamma}} > \frac{\Phi^{\frac{1}{\gamma}}}{h} \sqrt{\frac{m}{C}} \left[2\left(\frac{\pi}{3\sqrt{3}}\right)^{\frac{1}{3}} - 1\right] \tag{35}$$

Therefore if $\gamma > 2$ we get

$$\varepsilon < \varepsilon_M = \left(\frac{h^2 C}{m}\right)^{\frac{\gamma}{\gamma-2}} \frac{1}{\Phi^{\frac{2}{\gamma-2}}\left[2\left(\frac{\pi}{3\sqrt{3}}\right)^{\frac{1}{3}} - 1\right]^{\frac{2\gamma}{\gamma-2}}} \qquad (36)$$

hence the energy of particles in BEC is smaller than a maximum value $\varepsilon_M$ given by (36).
If $\gamma = 1$ (which is the case of Coulomb potential) we get

$$\varepsilon > \varepsilon_m = \frac{\Phi^2}{h^2}\frac{m}{C}\left[2\left(\frac{\pi}{3\sqrt{3}}\right)^{\frac{1}{3}} - 1\right]^2 \qquad (37)$$

or the energy is higher than a minimum value $\varepsilon_m$ given by (37).
In the next section we derive an expression for the interatomic potential (11), which is suited for neutral alkali-metal atoms and helium. The resulting value for $\gamma$ is $\gamma = 5$; therefore the "direction" of BEC states is toward the low energy region. On the contrary in the case of BEC of charged bosons, interacting through Colounb potential, obviously it is $\gamma = 1$ and only high energy BEC is allowed. The differences obtained in the energy is considerable: the calculations in the next two sections show $\varepsilon_M$ varies from $3.0 \times 10^{-15}$ erg to $3.0 \times 10^{-20}$ erg in the case of "cold" BEC of $^4$He and $^{87}$Rb respectively, depending on atomic dimensions, while $\varepsilon_m$ is about 0.01 Mev ($1.60 \times 10^{-8}$ erg) in the case of "hot" BEC of charged bosons (deuterons).

**4. Low energy BEC: the case of helium and alkali-metal atoms**
In this section we propose a method to calculate the interaction potential $\varphi(r)$ for neutral alkali-metal atoms and helium, which is consistent with known atomic and electron structure of these elements. The principal aim of the section is to determine exactly the constants $\Phi$ and $\gamma$ in order to give a numerical evaluation of the functions derived in Section 3.
In previous works on the subject based on variational and numerical methods, the Lennard-Jones potential is generally assumed as interaction potential (see for example [18])

$$\varphi(r) \propto \left[\left(\frac{\sigma}{r}\right)^{12} - \left(\frac{\sigma}{r}\right)^{6}\right]$$

since it is able to represent attractive and repulsive interactions, depending on the ratio $\sigma/r$. At low temperature the effect of the interaction potential in a two-body collision, can be described by the s-wave scattering length $a$ [14]. This effect depends on the sign of $a$: if $a>0$ the potential is repulsive, on the contrary if $a<0$ the potential is attractive. In the case of $^{23}Na$ and $^{87}Rb$ a repulsive potential with positive $a$ is generally accepted [27, 28], while in the case of $^7Li$ the literature reports different evaluations of the sign of the scattering length, depending on experimental conditions and hyperfine states involved in the two-body collisions [29, 30, 31]. Since we are interested only in determining the effects of interatomic interactions on the density function of quantum states (12), hence in the volume reduction due to such interactions, we will assume these ones as repulsive and caused by electrostatic charges of electrons and nuclei. Nowadays, after sodium, rubidium and lithium, all alkali elements [32, 33], hydrogen [34], metastable helium [35, 36], ytterbium [37] and chromium [38] have been added in the Bose-Einstein condensation list. It is noteworthy that the outermost electron orbital of all the elements cited above is an s-type (spherical) electron orbital, with one or two electrons. In the following this feature will be used to derive a simple expression for the interatomic potential in the case of such atoms. In order to calculate the interatomic potential (11) due to electrostatic interactions, we will consider only the effects of negative electronic charges in s-type orbitals and of the corresponding positive charges in the nuclei. Obviously these ones are all the existing charges in the case of helium, while in the case of alkali atoms we will assume the effects of the electrons in the inner orbitals and of the corresponding protons in the nuclei are negligible. Therefore the atomic model we adopt to calculate the electrostatic potential is composed of a positive point charge $+xe$ in its nucleus and a negative point charge $-xe$ at any position on a sphere with centre in the nucleus and radius $d$ (the s-type outermost orbital). As for the value of $x$ it is obviously $x=2$ in the case of helium and $x=1$ in the case of alkali-metal atoms. We use the concept of atomic radius as was first pointed out by Slater [39]: the distance from the nucleus of the principal maximum of the radial charge density distribution function of the outermost orbital. An updated discussion of the subject together with calculations of absolute radii of atoms and ions is given in [40]. Assuming the atomic model we have described above, if $r$ is the distance between two nuclei of interacting atoms, $e$ is the elementary charge in electrostatic units of cgs system ($e=4.80320 \times 10^{-10}$), it is easy to calculate the mean interatomic potential as

$$\varphi(r) = \frac{x^2 e^2}{r}\left(1 - \frac{2}{1+\frac{d^2}{3r^2}} + \frac{1}{1+\frac{2}{3}\frac{d^2}{r^2}}\right) \qquad (38)$$

where $r(1+ d^2/3r^2)$ is the average distance between an electron of the s-type orbital of the first atom and the nucleus of the second atom, $r(1+ 2d^2/3r^2)$ is the average distance between two electrons belonging to different s-type orbitals.

Since $(d^2/r^2) \ll 1$, by the approximate formula $(1+q)^{-1} \approx 1 - q + q^2$, which is valid for $|q| \ll 1$, with $q = d^2/3r^2$ and $q = 2d^2/3r^2$, from (38) we finally get

$$\varphi(r) = \frac{2}{9}\frac{d^4}{r^5}x^2 e^2 . \qquad (39)$$

Comparing (39) with (11), in the case of neutral atoms with outermost s-type orbital containing $x$ electrons, we get the values $\gamma = 5$ and

$$\Phi = \frac{2}{9}d^4 x^2 e^2. \qquad (40)$$

Since the exponent is $\gamma > 2$, in the case of $^4He$ and other bosonic alkali-metal atoms, Bose-Einstein condensation occurs at low energy level $\varepsilon < \varepsilon_M$ with $\varepsilon_M$ given by (36).

We can now try a numerical evaluation of functions (33) and (34) obtaining by this way the relation $T = T(n_\delta)$ between the temperature and the number density of the condensate. As for the atomic radii, we use the values of table 1, which are drawn from [40].

Table 1. Absolute Atomic Radii (see [40]).

| Atom | Atomic Radii $d$ ($\mathring{A}$) |
|---|---|
| Li-7 | 1.6282 |
| Na-23 | 2.1649 |
| Rb-87 | 4.8106 |
| He-4 | 0.3113 |

The results of this procedure are given in the following figures 3-7, where the graph of the function $T = T(n_\delta)$ is plotted for variuos atomic species and taking care of both energy and velocity modes (28) and (29), which correspond with parameters C2 and C1 defined in (31) and (32) respectively. In figures 3 and 4 the graph of the function is given on the whole scale of density in the case of $^4He$ and $^{87}Rb$, while in figures 5-7 the graph is plotted only for low density values and in the case of alkali-metal atoms.

In order to give a numerical example, the calculated values of some parameters interesting BEC are reported in table 2. They are evaluated in points $(n_\delta, T)$ chosen of the same order of the ones reported in current literature (besides the already mentioned references [1-3], see [5] and [41] for a survey of experimental work and observations and [42] for up-to-date and accurate measurements of the critical temperature $T$ of a dilute gas of $^{87}Rb$ atoms), and with the known lambda-point of $^4He$ (density about $1.88 \times 10^{22}$ cm$^{-3}$ at a temperature of 2.17 K).

**Table 2. Calculated values of BEC parameters for bosonic alkali atoms and He-4.**

| Atom (mode) | $\rho$ | $a$ ($\overset{\circ}{A}$) | $n_\delta a^3$ | $n_\delta$ (cm$^{-3}$) | $T$ (K) | $\varepsilon$ (erg) | $\varepsilon_M$ (erg) |
|---|---|---|---|---|---|---|---|
| Li-7 (C1) | 0.999 | 270 | $4.13 \times 10^{-5}$ | $2.10 \times 10^{12}$ | $3.62 \times 10^{-7}$ | $2.50 \times 10^{-23}$ | $1.46 \times 10^{-20}$ |
| Na-23 (C1) | 0.991 | 237 | $1.85 \times 10^{-3}$ | $1.39 \times 10^{14}$ | $2.15 \times 10^{-6}$ | $1.50 \times 10^{-22}$ | $4.32 \times 10^{-19}$ |
| Rb-87 (C2) | 0.991 | 772 | $1.78 \times 10^{-3}$ | $3.86 \times 10^{12}$ | $1.43 \times 10^{-7}$ | $1.00 \times 10^{-23}$ | $3.03 \times 10^{-20}$ |
| He-4 (C2) | 0.057 | 2.38 | 0.222 | $1.64 \times 10^{22}$ | 2.07 | $2.50 \times 10^{-15}$ | $3.01 \times 10^{-15}$ |

In the column of atomic species the velocity mode which best fits to the experimental data is also indicated (C2 is for (28) and C1 for (29)). The statistical theory here proposed does not allow us to decide what kind of mode is realized in the BEC of alkali-metal atoms in the low density range, while in the case of $^4He$ the only possible mode is (28) or C2, since the maximum energy in the case C1, $\varepsilon_M = 5.56 \times 10^{-16}$ erg as follows from (36), is lower than the energy calculated for superfluid helium. Note that in the case of alkali atoms the value of $\rho \approx 1$ is very different from the one for helium, $\rho \approx 0$, since in the former case the dilute gas hypothesis holds ($n_\delta a^3 \ll 1$, $n_0 \approx 0$, remember (22)), while in the latter one we have the case of a high density fluid ($n_\delta a^3 = 0.222$, $n_0 \approx 1$).

The values obtained in the case of superfluid $^4He$, $a = 2.38 \overset{\circ}{A}$, $n_\delta a^3 = 0.222$, are in good agreement with those reported in the studies on the subject, based on Monte Carlo simulations [43]. These values with the one of the corresponding energy, $\varepsilon = 2.50 \times 10^{-15}$ erg or $\varepsilon_H = 8.47$ in units $\hbar^2/ma^2$, are very close to data obtained in [18] as result of the Monte Carlo integration of the Schrodinger equation, for a system of 256 particles and phonon correction in order to consider an infinite system ($n_\delta a^3 = 0.244$, $\varepsilon_H = 8.50$).

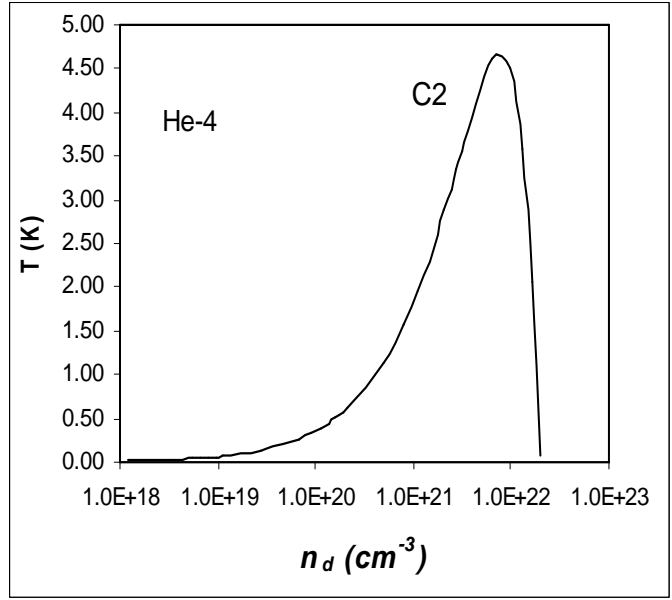

Figure 3 – BEC Temperature/Density functions for $^4$He. The whole range of admissible temperature and density values is plotted.

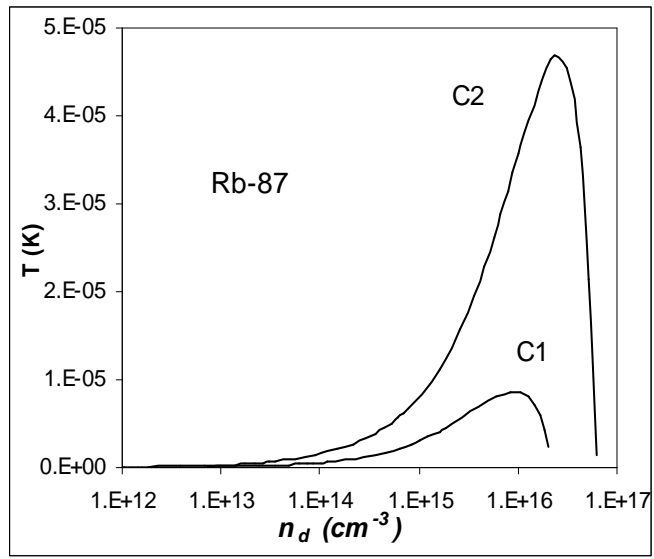

Figure 4 – BEC Temperature/Density functions for $^{87}$Rb. The whole range of admissible temperature and density values is plotted.

In the following figures we report the graphs of the equation of state for three atomic species in the low density region, which is the most experimentally investigated. A comparison with precise experimental data for this range of density (see [42] for rubidium), is outside the aim of this article and it will be presented in a subsequent work on the subject.

In the end we want to note that assuming the accepted values given in table 1 for atomic radii of different species, the calculated values of temperature and density in BEC obtained from the theory, are in agreement with those observed. Hence the theory gives an independent estimate of atomic radii.

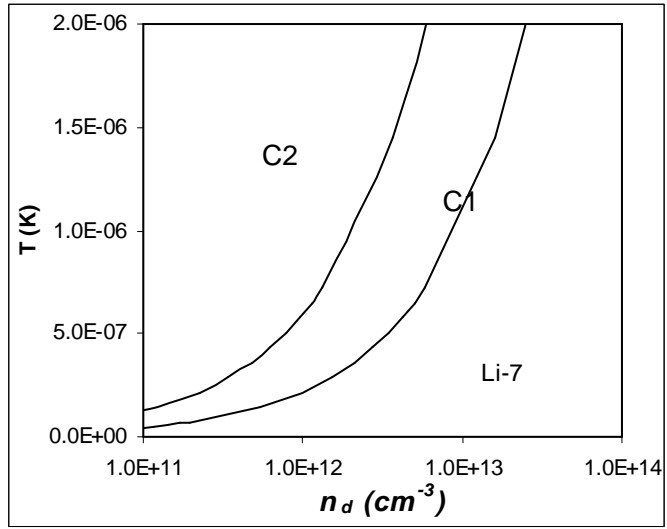

Figure 5 – BEC Temperature/Density functions for $^7$Li at low density.

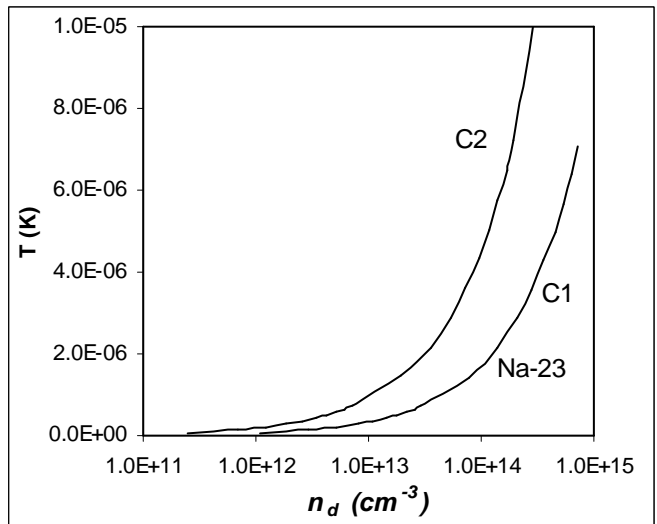

Figure 6 – BEC Temperature/Density functions for $^{23}$Na at low density.

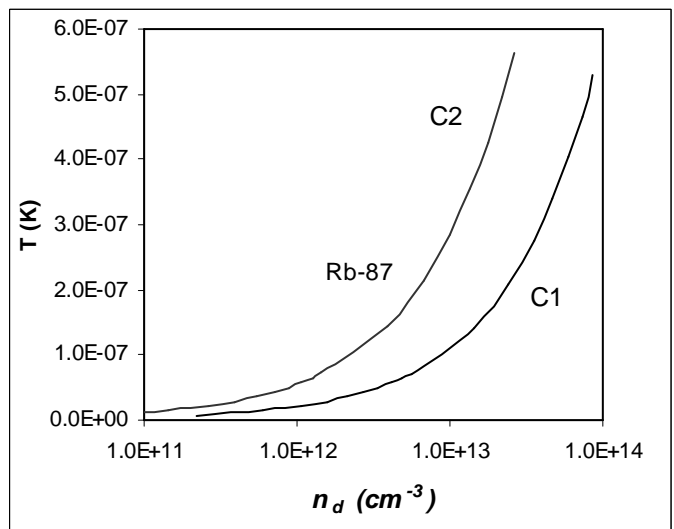

Figure 7 – BEC Temperature/Density functions for $^{87}$Rb at low density.

## 5. High energy BEC: the case of charged bosons in nuclear matter and the role of binding energy

In the following we consider a system of identical charged bosons in vacuum interacting through a central Coulomb potential

$$\varphi(r) = \frac{1}{r} x^2 e^2 \qquad (41)$$

($e$ is the elementary charge in electrostatic units of cgs system and $xe$, $x$ integer, is the positive point charge of the boson), hence the values of $\gamma$ in (11) and $w$ in (17) are given by $\gamma = 1$ and $w=5$. Adopting these definitions, from the generalized theory proposed in Section 2 and 3 in the case of deuterons ($x=1$), we derive the values $\varepsilon_m = 0.0105$ and $\varepsilon_m = 0.0290$ MeV for the minimum energy and velocity mode (28) and (29) respectively. In order to calculate the equation of state of deuterons in BEC, we have to take into consideration another important assumption about energy. When the two body interaction between bosons is of Coulomb type, it is known from equation (37) that only high energy BEC is possible. In other words the kinetic energy of interacting particles, must be large enough to win the repulsive Coulomb barrier and to get the right density value in order that BEC can occur. On the other hand the binding energy which is responsible for holding them tightly together in BEC state, must be equivalently large and the mass reduction due to this binding energy is no longer negligible: we have then assumed the binding energy between deuterons is equal to the energy $\varepsilon$ of particles in equations (33) and (34) (this assumption will be justified below in Appendix A). Therefore in performing the calculations about deuterons in BEC, according to the relativistic equation $E=mc^2$, stating the equivalence between energy $E$ and mass $m$ ($c$ is the velocity of light), we have to consider a reduced mass $m = m_0 - \varepsilon/c^2$, where $m_0$ is the rest mass of deuteron, in equations (33) and (34).

The resulting equation of state of deuterons in BEC is plotted in figure 8, for the whole range of admissible density and temperature values. The labels C2 and C1 still indicate the data obtained from equations (33) and (34) when the parameter C is given by (31) and (32) respectively. In abscissa the number density of nucleons (protons and neutrons) is reported, which is obtained simply by multiplying by 2 the number density of deuterons given by (33). Note that energy varies between $\varepsilon_m \leq \varepsilon < m_0 c^2$ and when $\varepsilon = m_0 c^2$, the limiting values for density and temperature are $n_\delta = 0$, $T = 2 m_0 c^2 /k$, the parameter $\rho$ defined in (18) is $\rho = 1$, as stated by (20) and (21). The BEC equation of state for deuterons in figure 8 describes the trajectory of the condensate through the phases of nuclear and hadronic matter (commonly referred to as nuclear equation of state), as presently depicted in current theories [44]. It doesn't mean that all the states of deuterons in BEC can exist, in particular the states above the deconfinement region between hadronic matter and quark-gluon plasma, which, according to [45, 46], is drawn starting at the temperature of 165–195 MeV for low baryon number content (or zero chemical potential) and then decreasing when the density becomes higher. The aim of figure 8 is to point out the important features of the equation of state: when energy varies in $[\varepsilon_m, m_0 c^2)$ temperature and nucleon density remain finite, ranging from $T = 0$ K, $n_\delta = 1.853 \times 10^{32} \div 3.886 \times 10^{33}$ cm$^{-3}$ (for C2 and C1 mode respectively) to the limiting values $T = 3756$ Mev (or $4.35 \times 10^{13}$ K), $n_\delta = 0$ as $\varepsilon$ approaches $m_0 c^2$. Density reaches its maximum value about $2.54 \times 10^{40} \div 1.15 \times 10^{41}$ cm$^{-3}$ when $T = 1881$ Mev ($2.18 \times 10^{13}$ K).

The states at low temperature $T = 0 \div 500$ K and minimum energy $\varepsilon_m$, corresponding with a value of $\rho \approx 0$, have a nucleon number density practically constant equal to $n_\delta = 1.853 \times 10^{32}$

cm$^{-3}$ for C2 mode ($n_\delta$ = 3.887 x10$^{33}$ cm$^{-3}$ for C1), which is the lowest value compatible with minimum energy BEC of bare deuterons in vacuum[2].

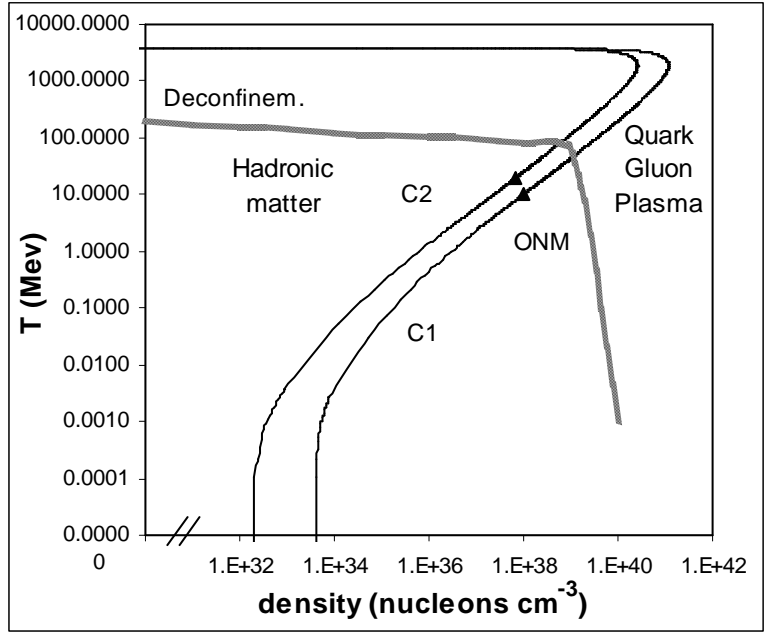

Figure 8 – BEC Temperature/Density functions for deuterons. In abscissa the number density of nucleons is reported (2 times the density of deuterons). The ONM label points out the ordinary nuclear matter region.

As we have said in the Introduction, Bose-Einstein condensation is now widely studied also as far as ordinary nuclear matter is concerned (see references in Section 1). The region corresponding with the ordinary nuclear matter (density about 10$^{38}$ nucleons cm$^{-3}$), is pointed out by the label ONM in figure 8. In the simple high energy BEC model we present in this work, stable nuclei are considered as built from deuterons $^2$H, assuming the particle energy $\varepsilon$ is equal to the binding energy of each interacting deuteron too, and $b=\varepsilon/2$ is the binding energy per nucleon usually considered in the literature about the subject. In figure 9 the relation between density and binding energy $b$, given by the BEC model based upon equation (33), is plotted (C2 parameter (31) is used). Nuclear density is expressed through the parameter $r_0$ which is used to calculate the nuclear radius as $R = r_0 A^{1/3}$ (A is the mass number). Strictly speaking, the model is suitable for nuclei having the same numer of protons and neutrons only: the points corresponding with some nuclei of this type are plotted in figure 9 (see [47] for the binding energy $b$). It is noteworthy that, in despite of model simplicity, the values obtained for $r_0$ are in good agreement with those derived from scattering experiments [48], therefore the method is suitable to give an independent theoretical estimate of nuclear radii.

---

[2] This value is strongly dependent on the constant $\Phi = e^2$ in (41). Assuming the deuterons are in a medium with some electron screening effect of the Coulomb potential, hence with constant $\Phi << e^2$ in (41), the equation of state predicts much lower values for the number density in low temperature BEC states. These are consistent with the ones observed in some experiments with low energy reactions of deuterons in metal targets (see [49, 50] and references therein), where an enhancement of the reaction cross-section is registered with decreasing energy.

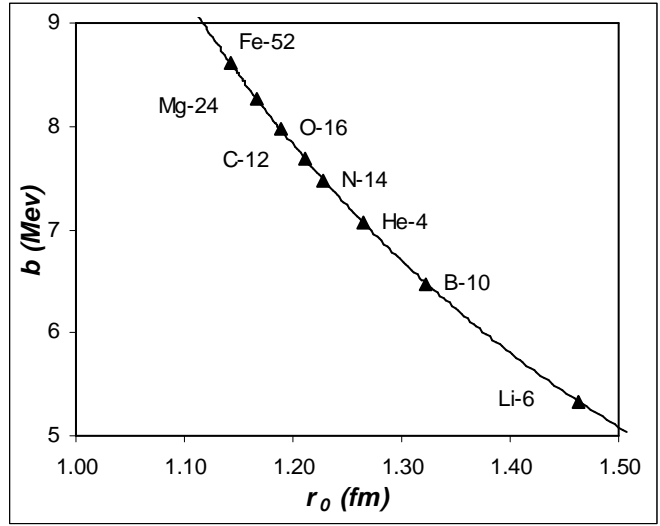

Figure 9 – Binding energy/Density function for BEC model of stable nuclei built from deuterons. Density is expressed through $r_0$, linking nuclear radius R and mass number A ($R=r_0 A^{1/3}$). The points corresponding with some nuclei having the same number of protons and neutrons are reported.

## 6. Summary and conclusions

Given a system of identical bosons interacting each other through a central potential of the type $\varphi(r) \propto 1/r^\gamma$, we derived the general equation of state for the system in Bose-Einstein condensation, assumed it is in its ground state, and demonstrate that the energy (and temperature) of particles in BEC, depends on the value of $\gamma$. If $\gamma > 2$ only low energy/temperature BEC is allowed; on the contrary if $\gamma = 1$ (which is the case of charged bosons), only high energy/temperature BEC can occur. Numerical results from the equation of state are given, in the case of $^4$He and alkali metals atoms, for which a suitable model of two-body interaction potential is proposed with $\gamma = 5$, involving the atomic radii too, and in the case of charged deuterons in stable nuclei. In this case the effect of binding energy between bosons in BEC is not negligible, and it is considered in connection with the energy of particles and nuclear dimensions. The theory gives an independent theoretical estimate of atomic and nuclear radii depending on temperature and density of BEC.

## Appendix A

In the following we briefly investigate the conditions in order that the assumption, we have made in Section 5, about the equality of the binding energy and the kinetic energy $\varepsilon$ of deuterons within the nucleus, can hold.

Let us consider a nucleus in its ground state, with atomic number Z and mass number $A=2Z$. The nucleus can be thought as composed of Z deuterons D (or $^2$H) interacting each other, and we assume a central square-well crater like interaction potential, between a single boson D and the remaining nucleus with charge (Z-1):

$$V(r) = -V_0 \quad \text{as } 0 \leq r < R \text{ (with } V_0>0\text{)}$$

$$V(r) = \Phi_Z / r \quad \text{as } r \geq R \text{ (with } \Phi_Z = (Z-1)e^2\text{)}$$

(r is the radial distance between the boson D and the centre of the nucleus).

Since $V_0>0$, the interaction is attractive within the nucleus (R can be assumed as the nuclear radius) and repulsive outside.

Given $\psi(r) = u(r)/r$ the radial part of the wave function of the boson D, the time-independent Schrodinger equation for the ground state of the boson is

$$\frac{d^2u(r)}{dr^2} - \frac{2m}{\hbar^2}[V(r) - E]u(r) = 0$$

where $\hbar = h/2\pi$ and $E$ is the total energy of the particle, which can be written also as

$$E = -V_0 + \varepsilon.$$

Since we are interested only in writing the boundary conditions of the wave functions in $r = R$, we can assume a stepwise potential function $V(r)$ around $R$, or

$$V(r) = \Phi_Z / R \quad \text{when } R \leq r < R + \delta$$

with $\delta > 0$ and small. Then, looking for a solution of the type

$$u(r) = \exp(y(r)/\hbar)$$

under the assumptions $\psi(r)$ is defined in $r=0$ and quadratically integrable as $r \to \infty$, it is easy to get

$$u(r) = u_1(r) = A \sin\left(\frac{\sqrt{2m\varepsilon}}{\hbar} r\right) \quad \text{as } 0 \leq r < R$$

and

$$u(r) = u_2(r) = B \exp\left(-\frac{\sqrt{2m\beta}}{\hbar}(r - R)\right) \quad \text{as } R \leq r < R + \delta$$

with

$$\beta = \frac{\Phi_Z}{R} - E.$$

Note that $\beta$ is the binding energy of the boson D, since it is the difference between the top of the potential barrier $V(R)$ and the total energy of the particle.

In order that the wave function is continuous in $r = R$, the following conditions are to be satisfied

$$u_1(R) = u_2(R)$$
$$\frac{du_1(r)}{dr} = \frac{du_2(r)}{dr} \quad \text{as } r=R$$

from which equations the relation between the kinetic energy $\varepsilon$ and the binding energy $\beta$ of the particle follows (continuity condition):

$$\sqrt{\varepsilon} \cot\left(\frac{\sqrt{2m\varepsilon}}{\hbar} R\right) = -\sqrt{\beta}.$$

Let us indicate with

$$\lambda = \frac{h}{\sqrt{2m\varepsilon}}$$

the De Broglie wavelength of the particle D within the nucleus, then the above equation can be rewritten as

$$\sqrt{\varepsilon} \cot\left(\frac{2\pi R}{\lambda}\right) = -\sqrt{\beta}.$$

Put

$$\frac{2R}{\lambda} = q + k, \quad k \text{ integer,}$$

then in order that the continuity condition is satisfied, $q$ must be $1/2 < q < 1$ and it is $\varepsilon = \beta$ if $q = 3/4$.

**References**

[1] Anderson M H, Ensher J R, Matthews M R, Wieman C E and Cornell E A 1995 Observation of Bose-Einstein condensation in a dilute atomic vapor *Science* **269** 198
[2] Bradley C C, Sackett C A and Hulet R G 1997 Bose-Einstein condensation of lithium: observation of limited condensate number *Phys. Rev. Lett.* **78** 985
[3] Davis K B, Mewes M-O, Andrews M R, van Druten N J, Durfee D S, Kurn D M and Ketterle W 1995 Bose-Einstein condensation in a gas of sodium atoms *Phys. Rev.Lett.* **75** 3969
[4] Dalfovo F, Giorgini S, Pitaevskii L P and Stringari S 1999 Theory of Bose-Einstein condensation in trapped gases *Rev. Mod. Phys.* **71** 463
[5] Ketterle W 2002 Nobel lecture: when atoms behave as waves: Bose-Einstein condensation and the atomic laser *Rev. Mod. Phys.* **74** 1131
[6] Griffin A, Snoke DW and Stringari S 1995 *Bose-Einstein condensation* (Cambridge: Cambridge University Press)
[7] Kim Y E and Zubarev A L 1998 *Proc. of the 7th Int. Conf. on Cold Fusion (ICCF-7) (Vancouver Canada19-24 April1998)* pp 186-191
[8] Kim Y E and Zubarev A L 2001 Ground state of charged bosons confined in a harmonic trap *Phys. Rev. A* **64** 013603
[9] Kim Y E 2004 Quantum many-body theory and mechanisms for low energy nuclear reaction processes in matter *Proc. of the Int. Conf. Fusion 03: From a Tunneling Nuclear Microscope to Nuclear Processes in Matter (Matsushima Miyagi Japan 12-15 November 2003) Prog. Theor. Phys. Supplement* **154**
[10] Tohsaki A, Horiuchi H, Schuck P and Ropke G 2001 Alpha cluster condensation in $^{12}$C and $^{16}$O *Phys. Rev. Lett.* **87** 192501
[11] Yamada T and Schuck P 2004 Dilute multi-alpha cluster states in nuclei *Phys. Rev. C* **69** 024309
[12] Gridnev K A, Torilov S Yu, Kartavenko V G and Greiner W 2007 Model of binding alpha-particles and structure of the light nuclei *Int. J. Mod. Phys. E* **16** 1059
[13] Schuck P, Funaki Y, Horiuchi H, Ropke G, Tohsaki A and Yamada T 2007 Quartetting in fermion matter and alpha particle condensation in nuclear systems *Romanian Reports in Physics* **59** 675
[14] Huang K 1987 *Statistical Mechanics* (New York: John Wiley and Sons)
[15] Barbarani V 2007 New quantum statistics and the theory of Bose-Einstein condensation *Int. J. Theor. Phys.* **46** 2401
[16] Hansen J P, Levesque D and Shiff D 1971 Fluid-Solid Phase Transition of a Hard-Sphere Bose System *Phys. Rev. A* **3** 776
[17] Huang K 1964 *Studies in Statistical Mechanics* vol. II (Amsterdam: North-Holland)
[18] Kalos M H, Levesque D and Verlet L 1974 Helium at zero temperature with hard-sphere and other forces *Phys. Rev. A* **9** 2178
[19] Lee T D and Yang C N 1957 Many-Body problem in quantum mechanics and quantum statistical mechanics *Phys. Rev.* **105** 1119
[20] Lee T D Huang K and Yang C N 1957 Eigenvalues and eigenfunctions of a Bose system of hard spheres and its low-temperature properties *Phys. Rev.* **106** 1135
[21] Oden L, Henderson D and Coleman J 1964 Quantum cell model for hard spheres *Proc. Nat. Acad. Sci. USA* **51** 629
[22] Barker J A 1960 A new theory of fluids: the 'tunnel' model *Aust. J. Chem.* **13** 187
[23] Born M 1969 *Atomic Physics* (Glasgow: Blackie)
[24] Kompaneyets A S 1961 *Theoretical Physics* English edition (Moscow: Peace Publishers) p 445
[25] Solis M A, de Llano M, Clark J W and Baker G A Jr 2007 Improved quantum hard-sphere ground-state equation of state *Phys. Rev. E* **76** 031125
[26] Henderson D 1971 Tunnel model for hard-sphere bosons in their ground state *Proc. Nat. Acad. Sci. USA* **68** 2011
[27] Tiesinga E, Williams C J, Julienne P S, Jones K M, Lett P D and Phillips W D 1996 A spectroscopic determination of scattering lengths for sodium atom collisions *J. Res. Natl Inst. Stand. Technol.* **101** 505
[28] Gardner J R, Cline R A, Miller J D, Heinzen D J, Boesten H M J M and Verhaar B J 1995 Collisions of doubly spin-polarized ultracold $^{85}$Rb atoms *Phys. Rev. Lett.* **74** 3764
[29] Abraham E R I, McAlexander W I, Sackett C A and Hulet R G 1995 Spectroscopic determination of the s-wave scattering length of Lithium *Phys. Rev. Lett.* **74** 1315
[30] Abraham E R I, McAlexander W I, Gerton J M, Hulet R G, Coté R and Dalgarno A 1996 Singlet s-wave scattering lengths of $^{6}$Li and $^{7}$Li *Phys. Rev. A* **53** R3713
[31] Moerdijk A J and Verhaar B J 1994 Prospects for Bose-Einstein condensation in atomic $^{7}$Li and $^{23}$Na *Phys. Rev. Lett.* **73** 518

[32] Modugno G, Ferrari G, Roati G, Brecha R J, Simoni A and Inguscio M 2001 Bose-Einstein condensation of potassium atoms by sympathetic cooling *Science* **294** 1320
[33] Weber T, Herbig J, Mark M, Nagerl H C and Grimm R 2003 Bose-Einstein condensation of cesium *Science* **299** 232
[34] Fried D G, Killian T C, Willmann L, Landhuis D, Moss S C, Kleppner D and Greytak T J 1998 Bose-Einstein condensation of atomic hydrogen *Phys. Rev. Lett.* **81** 3811
[35] Pereira Dos Santos F, Lonard J, Junmin W, Barrelet C J, Perales F, Rasel E, Unnikrishnan C S, Leduc M and Cohen-Tannoudji C 2001 Bose-Einstein condensation of metastable helium *Phys. Rev. Lett.* **86** 3459
[36] Robert A, Sirjean O, Browaeys A, Poupard J, Nowak S, Boiron D, Westbrook C I and Aspect A 2001 A Bose-Einstein condensate of metastable atoms *Science* **292** 461
[37] Takasu Y, Maki K, Komori K, Takano T, Honda K, Kumakura M, Yabuzaki T and Takahashi Y 2003 Spin-singlet Bose-Einstein condensation of two-electron Atoms *Phys.Rev. Lett.* **91** 040404
[38] Griesmaier A, Werner J, Hensler S, Stuhler J and Pfau T 2005 Bose-Einstein condensation of chromium *Phys. Rev. Lett.* **94** 160401
[39] Slater J C 1964 Atomic radii in crystals *J. Chem. Phys.* **41** 3199
[40] Dulal C G and Raka B 2002 Theoretical calculation of absolute radii of atoms and ions. Part 1. The atomic radii *Int. J. Mol. Sci.* **3** 87
[41] Cornell E A, Ensher J R and Wieman C E 1999 Experiments in dilute atomic Bose-Einstein condensation *Bose-Einstein Condensation in Atomic Gases. Proc. of the Intern. School of Phys. Enrico Fermi. Course CXL* Inguscio M, Stringari S and Wieman C E (Italian Physical Society)
[42] Gerbier F, Thywissen H, Richard S, Hugbart M, Bouyer P and Aspect A 2004 Critical temperature of a trapped weakly interacting Bose gas *Phys. Rev. Lett.* **92** 030405
[43] Gruter P, Ceperley D and Laloe F 1997 Critical temperature of Bose-Einstein condensation of hard-sphere gases *Phys. Rev. Lett.* **79** 3549
[44] Bertulani C A and Danielewicz P 2004 *Introductions to Nuclear Reactions* (Bristol, UK : IOP Publishing) pp 506-509
[45] Danielewicz P 2001 Nuclear equation of state *Preprint* arXiv:nucl-th/0112006v1
[46] Aichelin J and Schaffner-Bielich J 2008 The quest for the nuclear equation of state *Preprint* arXiv:0812.1341v2
[47] Audi G and Wapstra A H 1995 *Nucl. Phys.* **A595** 409
[48] Krane K S 1988 *Introductory Nuclear Physics* (USA : John Wiley and Sons Inc.) pp 45-58
[49] Raiola F, Gang L, Bonomo C, Gyurky G, Aliotta M, Becker H W, Bonetti R, Broggini C, Corvisiero P, D'Onofrio A, Fulop Z, Gervino G, Gialanella L, Junker M, Prati P, et al 2004 Enhanced electron screening in d(d,p)t for deuterated Ta *Eur. Phys. J. A* **19** 283
[50] Kim Y E and Zubarev A L 2007 Theoretical interpretation of anomalous enhancement of nuclear reaction rates observed at low energies with metal targets *Jpn. J. Appl. Phys.* **46** 1656-1662